\documentclass{ifacconf}

\usepackage{mathrsfs,amsmath,amssymb,amsfonts,mathtools}
\usepackage{natbib}
\usepackage{xcolor}
\usepackage{bm}
\usepackage{amsfonts}
\usepackage{listings} 
\usepackage{graphicx}  
\usepackage{subcaption}

\newtheorem{remark}{Remark}
\newtheorem{example}{Example}

\begin{document}
\begin{frontmatter}

\title{Biomolecular LQR under Partial Observation. \thanksref{footnoteinfo}} 

\thanks[footnoteinfo]{This work was funded by the National Natural Science Foundation of China under Grant No. 62303409 and 12320101001, the Jiangsu Provincial Scientific Research Center of Applied Mathematics under Grant No. BK20233002.}

\author[First]{Xiaoyu Zhang} 
\author[Third]{Zhou Fang}
\address[First]{ School of Mathematics, Southeast University, Nanjing, 211189, China(e-mail: xiaoyu\_z@seu.edu.cn).}
\address[Third]{State Key Laboratory of Mathematical Sciences, Academy of Mathematics and Systems Science, Chinese Academy of
Sciences, Beijing, 100190, China. (e-mail: zhfang@amss.ac.cn).} 

\begin{abstract}                
 This paper introduces a biomolecular Linear Quadratic Regulator (LQR) to investigate the design principles of gene regulatory networks. We show that for fundamental gene regulation network, the bio-controller derived from LQR theory precisely recapitulate natural network motifs, such as auto-regulation and incoherent feedforward loops.  This emulation arises from a fundamental principle: the LQR cost function mathematically encodes environmental survival demands, which subsequently drives the selection of both network topology and biochemical parameters. Our work thus establishes a theoretical basis for interpreting biological circuit design, directly linking evolutionary pressures to observable regulatory structures.

\end{abstract}


\end{frontmatter}

\section{Introduction}
Regulation in biological systems refers to the capacity of cells to actively maintain internal variables, such as molecular concentrations, around specific target values, or set-points \citep{Aoki2019, Del2016}. This capability is essential for sustaining cellular function and organismal health. For instance, the concentration of blood glucose is controlled by the coordinated action of insulin and glucagon to remain near a physiological set-point, preventing both hypoglycemia and hyperglycemia 
\citep{Wo1995}. Similarly, certain regulatory proteins, such as p53, exhibit low basal expression levels under normal conditions and are transiently upregulated only in response to stress, after which they return to their original set-points \citep{KG2009}. These examples highlight the fundamental importance of set-point regulation in ensuring system stability and functionality \citep{Kh2018}. Inspired by such natural mechanisms, the construction of synthetic biological systems capable of autonomously regulating protein expression toward desired levels has become a key objective in the field of synthetic biology  \citep{Del2016}.

In recent years, feedback control has been increasingly applied to the design of synthetic biological circuits \citep{filo2023biomolecula}. A typical feedback system consists of a plant—the system to be regulated—and a controller that adjusts the input 
$u$ to drive the system toward a desired set point 
$\bar{x}_{\text{sp}}$
  \citep{PR1999}. In control theory, various control strategies have been developed to meet different objectives, such as proportional-integral-derivative (PID) controller, 
   linear quadratic regulation (LQR), robust control. The central challenge of its application in  synthetic biology lies in how to implement these control architectures using biologically feasible reaction networks, relying on transcriptional, translational, or enzymatic processes.

Owing to its simplicity and practical effectiveness, the PID controller has emerged as a successful and widely adopted approach in this interdisciplinary domain. Integral controller is the key mechanism for robust regulation \citep{A2019, Aoki2019}, which has been found in bacterial chemotaxis in E.coil \citep{Yi2000}, temperature control \citep{Ni2009}, regulation of osmolarity \citep{M2009}, etc. To further improve the performance of transient response and noise reduction, antithetic integral feedback combined with proportional feedback (PI controller) \citep{B2018} and molecular PID controllers of different degrees of complexity are designed \citep{C2019,F2022}. 

Beyond achieving accurate control, biological systems must also consider factors such as energy consumption, resource utilization, and response speed. Thus biological regulatory processes often involve trade-offs among multiple competing objectives, reflecting an inherent optimality in regulatory strategies \citep{TJ2022}. This balance is not necessarily the result of explicit design, but rather an outcome of evolution. For example, in calcium homeostasis, cells must respond rapidly to external stimuli while avoiding excessive energy expenditure or signal disruption—thus achieving a dynamic balance between response efficiency and regulatory cost \citep{B2003}. These observations motivate the modeling of biological regulation as an optimal control problem.

To address these trade-offs, we employ linear quadratic regulation (LQR), which enables the systematic design of controllers that balance regulation accuracy with control effort. This paper considers the classical one- and two-gene expression system as the underlying plants, and demonstrates the implementation of LQR control via a collection of simple  biochemical reactions. As LQR fundamentally relies on full-state feedback, accurate measurement and estimation of the system states are crucial for its successful implementation. To address the challenges posed by unmeasurable states in two classes of systems, we design reduced-order state observers and subsequently construct LQR-based closed-loop systems using the estimated states. Notably, some of designed closed-loop systems align with commonly occurring motif, such as negative autoregulation motif loops and incoherent feedforward loop (IFFL). The remainder of the paper is organized as follows. Section \ref{sec:2} presents the basic concepts about LQR. Section \ref{sec:3} designs the bimolecular LQR for the one and two gene expression process. Section \ref{sec:4} conduct some numerical simulation to present the effective of our design. Section \ref{sec:5} concludes the paper and give the future research directions.
\section{Preliminaries}\label{sec:2}
Single-input single-output (SISO) systems are favored for their simple structure and ease of implementation. This section reviews LQR for SISO bio-system which will be implemented by deterministic biochemical reaction networks in the subsequent sections. 

Assume that the dynamics of the
 deterministic SISO biochemical system can be captured by a set of ODEs
\begin{equation}\label{eq:de}\dot{x}=f(x)+u(t)e_1
\end{equation}
where $x\in\mathbb{R}^n_{\geq 0}$ is the concentration state vector and $u$ is the control input, $e_1=[1,0,\cdots,0]^{\top}\in\mathbb{R}^n_{\geq 0}$.
Denote  $\bar{x}$ as the desirable set point and we perform linear perturbation analysis of system \eqref{eq:de} around $\bar{x}$. The derived linearized system is a continuous time system 
\begin{equation}\label{eq:lsde}
\dot{\tilde{x}}=A\tilde{x}+B\tilde{u}.
\end{equation}
Here, $\tilde{x}(t)=x(t)-\bar{x}$, $\tilde{u}(t)=u(t)-\bar{u}$ where $\bar{u}$ satisfies $f(\bar{x})+\bar{u}e_1=0$. 

In control theory, LQR provides a design framework that achieves a trade-off between regulation performance and control cost by minimizing the cost functional
\begin{equation}\label{eq:cf}
J=\int_t^{+\infty}\tilde{x}(s)^{\top}Q\tilde{x}(s)+\tilde{u}^{\top}(s)R\tilde{u}(s)ds.
\end{equation}
Here, $Q=Q^{\top}\succeq 0$ and $R$ are all weighted matrix. Furthermore, for the SISO system described by \eqref{eq:de}, we assume $R=1$ without loss of generality.  
The corresponding optimal theoretical control is the state feedback law
$$
\tilde{ u}_{\text{To}}(t) = -K\tilde{x}(t),
$$
where $K= B^\top P$, with $P = P^\top \succeq 0$ solving the algebraic Riccati equation
\begin{equation}\label{eq:riccati}
A^\top P + P A - PB B^\top P + Q = 0.
\end{equation}

Finally, the original optimal control input for the linear perturbation system is recovered as
$$   u_{\text{To}} = \bar{u}-K (x(t) -\bar{x}).
$$ 

Although LQR theory provides a control law, $u_{\text{To}}$, that both balances control effort and tracking precision. However, such a structure form rarely  aligns with the types of interaction pathways found in biological systems. Therefore, in this work, we design the optimal bio-regulator $u_{\text{bo}}(t)$ to approximate the theoretical LQR controller $u_{\text{To}}(t)$. An observer is also designed to estimate the full state of the system from the partial observation.

\section{The LQR closed loop design}\label{sec:3}
\subsection{LQR design of one-gene expression process}\label{subsec:31}
\begin{figure}[h]	 \begin{subfigure}{.24\textwidth}		\centering  		\includegraphics[width=0.75\textwidth]{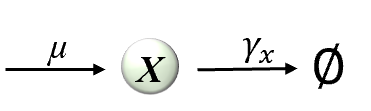}	 	\caption{open loop}\label{fig:1a}
   \end{subfigure}
     \begin{subfigure}{.24\textwidth}    \centering\includegraphics[width=0.75\textwidth]{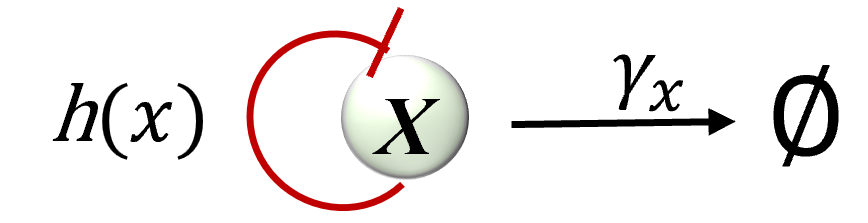}
      \caption{closed loop}\label{fig:1b}
     \end{subfigure}
\caption{LQR design of birth-death process}
     \label{fig.1}
\end{figure}

The one gene expression process known as birth-death process (Fig. \ref{fig:1a}) can be captured as follows:
\begin{equation}
\dot{x}=\mu-\gamma_xx
\end{equation}
where $\mu$ and $\gamma_x$ denote the synthesis and degradation rate constants of $X$, respectively. To regulate $X$ to a desired set point $\bar{x}_{\text{sp}}$, $\mu$ can be tuned as $\mu=\gamma_x\bar{x}_{\text{sp}}$. 

 The presence of the disturbance necessitates closed-loop regulation. To balance the regulation error and cost effort, we employ LQR theory to derive the  regulation law for this system. Given that the system has a single state $X$ which is both the controlled variable and the output, the state penalty weight $Q$ in the cost function $J$ is a positive scalar. Theoretically, the regulation law takes the form of state feedback $u_{\text{To}}=\mu-p_x(x-\bar{x}_{\text{sp}})$
 where $p_x=\sqrt{\gamma_x^2+Q}-\gamma_x$ satisfies 
\begin{equation}
-2\gamma_xp_x+Q-p_x^2=0.
\end{equation}


However, such a structure is often biologically implausible. Therefore, we implement the LQR using a common motif in gene regulatory network — a negative autoregulation loop shown in Figure~\ref{fig:1b} \citep{RN2002}. The dynamics of the controller is given by:
\begin{equation}\label{eq:d1}
\dot{x}=-\gamma_x x+u_{\text{bo}}(t)=-\gamma_x x+\kappa h(x).   
\end{equation}
Here, $h(x) \triangleq \frac{K_x}{K_x + x}$ denotes the Hill function. The corresponding linearized system around the equilibrium point $x = \bar{x}_{\text{sp}}$ can be expressed as
\begin{equation}\label{eq:cl2}
    \dot{x}=-\gamma_x x+\kappa h(\bar{x}_{\text{sp}})-\tilde{\kappa}(x-\bar{x}_{\text{sp}})
\end{equation}
where $\tilde{\kappa}$ is defined as $\tilde{\kappa}=\kappa\frac{K_x}{(K_x+\bar{x}_{\text{sp}})^2}$. 

Next, if $\kappa$ and $K_x$ satisfy the following conditions:
\begin{equation}
    \begin{split}
      \kappa h(\bar{x}_{\text{sp}})=\mu=\gamma_x\bar{x}_{\text{sp}}; ~\tilde{\kappa}=\sqrt{\gamma_x^2+Q}-\gamma_x,
    \end{split}
\end{equation}
the linearization of $u_{\text{bo}}(t)$ in   (\ref{eq:d1}) is exactly equivalent to $u_{\text{To}}$ given by LQR theory.
Namely, the parameters can be explicitly determined as $K_x= \bar{x}_{\mathrm{sp}} \left( \frac{\gamma_x}{\sqrt{\gamma_x^2 + Q} - \gamma_x} - 1 \right)$ and $\kappa = \frac{ \gamma_x^2 \, \bar{x}_{\mathrm{sp}} }{ 2\gamma_x - \sqrt{\gamma_x^2 + Q}}$, under which the negative autoregulation loop  functions exactly as an LQR that minimizes the given quadratic cost function.
\begin{remark}
The weight $Q$ embodies a trade-off between regulation error and control effort. A larger $Q$ signifies a higher priority on tracking precision, often at the expense of increased control cost. Crucially, since both parameters $K_x$ and $\kappa$ of the Hill function must be positive for a  negative feedback motif, our derived analytical solution imposes an inherent constraint: $Q < 3\gamma_x^2$. Actually, this is not merely a mathematical result. It reveals the maximum performance ceiling—the most aggressive control strategy—that can be physically implemented by the simple negative autoregulation motif. Namely, the maximum control effort that this specific biological circuit architecture can be designed to exert.
\end{remark}
\begin{remark}
In the absence of self-inhibition, the production of protein $X$ can be considered proportional to the concentration of its gene $G_X$, which remains constant. As a result, the production rate of $X$ can be modeled as a constant input. In contrast, when self-inhibition is present, the production of $X$ is no longer constant but instead follows a Hill-type regulatory function that decreases with increasing $X$ concentration.
\end{remark}
\begin{remark}
For each given ideal set point value $\bar{x}_{\text{sp}}$, the corresponding values of $K_x$ and $\kappa$ can be derived analytically. These parameters can be experimentally tuned: $\kappa$ can be adjusted by controlling the promoter strength, which affects the maximal production rate of protein $X$, while $K_x$ can be modulated by modifying the binding affinity of the regulatory elements that mediate self-inhibition. 
\end{remark}
\begin{remark}
 This controller functions not only as a proportional controller but also exhibits a low-pass filtering effect. This behavior arises because the self-inhibition, modeled by a Hill function, is effectively derived from a quasi-steady-state approximation of the fast binding and unbinding kinetics between protein $X$ and its promoter region. Consider the following biochemical reactions:
  \begin{equation*}
\underbrace{G_{X}\xrightarrow{\kappa}X+G_{X};G_{X}+X\xrightarrow{k_1}G_{X}X}_{\text{fast process}}; \underbrace{G_{X}X\xrightarrow{k_2} G_{X}+X}_{\text{slow process}}.
 \end{equation*}
 The first two reactions are fast reaction process and 
 the corresponding dynamics is
 \begin{equation}
    \dot{[G_{X}]}=-k_1x[G_{X}]+k_2[G_XX]
 \end{equation}
 and the slow dynamics is in the form of
 \begin{equation}
     \dot{x}=-\gamma x+k_1[G_{X}].
 \end{equation}
 By applying the Laplace transform under the assumption of quasi-steady-state for $[G_X]$, we obtain the transfer function of the controller input
$$u(s)=\kappa [G_{X}](s)=\kappa\frac{k_2G_T-k_1\bar{[G_X]}x(s)}{s+k_1\bar{x}_{\text{sp}}+k_2}$$
where $G_T$ denotes the total concentration of gene $X$. This expression reveals that the controller behaves like a first-order low-pass filter with cutoff frequency $\omega_c = k_1 \bar{x}_{\text{sp}} + k_2$. 
\end{remark}
\subsection{LQR design of the two-gene expression process}
\begin{figure}[h]	 \begin{subfigure}{.24\textwidth}		\centering  		\includegraphics[width=0.75\textwidth]{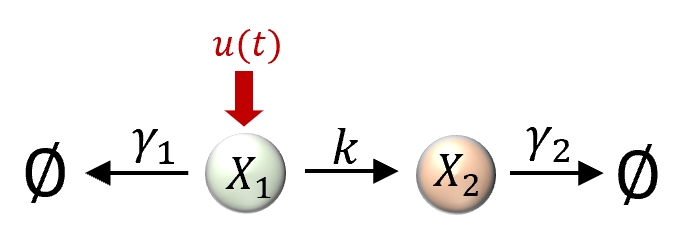}	 	\caption{open loop}\label{fig2a}
   \end{subfigure}
     \begin{subfigure}{.24\textwidth}    \centering\includegraphics[width=0.75\textwidth]{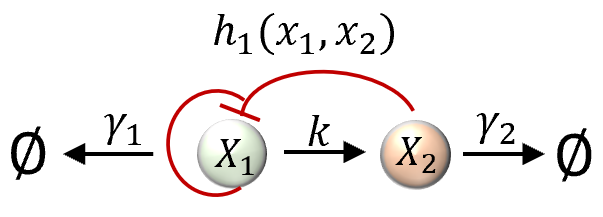}
      \caption{All states are measurable.}\label{fig2b}
     \end{subfigure}
      \begin{subfigure}{.24\textwidth}    \centering\includegraphics[width=0.75\textwidth]{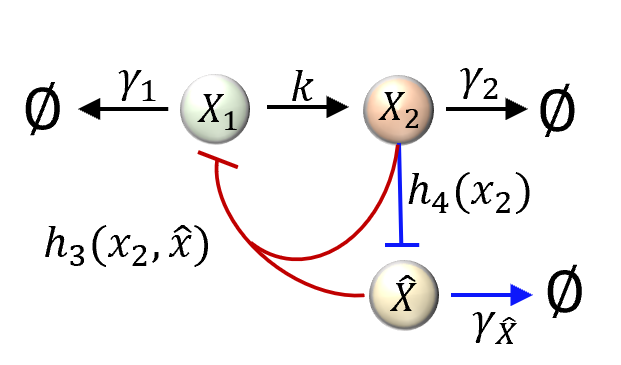}
      \caption{$X_2$ is measurable.}\label{fig2c}
     \end{subfigure}
     \begin{subfigure}{.24\textwidth}    \centering\includegraphics[width=0.79\textwidth]{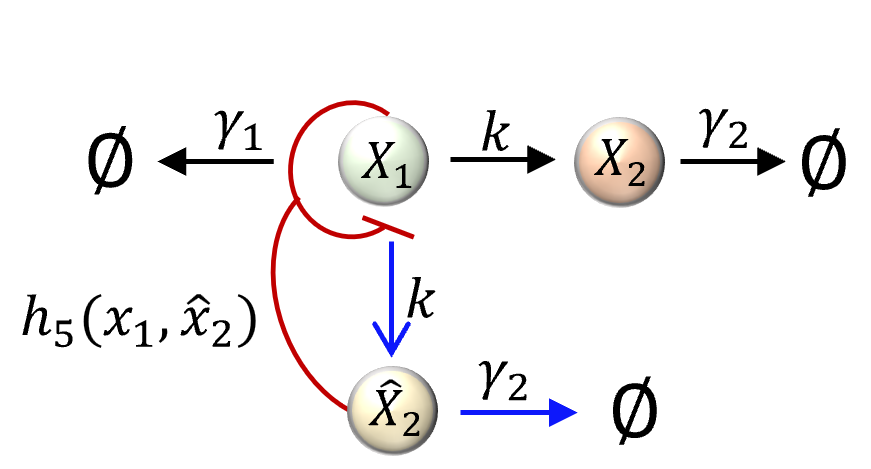}
      \caption{$X_1$ is measurable.}\label{fig2d}
     \end{subfigure}
\caption{LQR of Case I for two-gene expression process.}
     \label{fig.2}
\end{figure}

This subsection focuses on the bio-LQR design for a SISO two-gene expression system, in which the expression of gene $X_2$ is regulated by gene $X_1$ to some set-point $\bar{x}_{\text{sp}}$, as illustrated in Fig. \ref{fig2a}. In this configuration, the  output is defined as $X_2$, thus the weighting matrix $Q$ is constructed such that $Q_{22}$ is its only non-zero element. The controller design will be conducted separately for the cases where the control input is applied to $X_1$ and $X_2$, respectively.

\textbf{Case I: Actuation via species $X_1$.} We first focus on regulating the overall system through actuating species $X_1$.
The dynamics of the closed loop system is given by
\begin{equation}\label{eq:ds}
\begin{split}
\dot{x}_1=u(x)-\gamma_{1}x_1;~~\dot{x}_2=kx_1-\gamma_2x_2.
\end{split}
\end{equation}
Based on the formulation of Section \ref{sec:2}, the LQR controller takes the form  $u_{\text{To}}=-P_{11}(x_1-\bar{x}_1)-P_{12}(x_2-\bar{x}_{\text{sp}})+\bar{u}$ where $P_{ij}, i,j=1,2$ are the elements of the solution $P$ to the algebraic Riccati equation. Since the system is stabilizable, the solution $P$ is guaranteed to be positive definite.  Moreover, by solving the Riccati equation explicitly, the elements of $P$ satisfy 
\begin{equation}
P_{11}^2-2kP_{12}+2\gamma_1P_{11}=0.
\end{equation} 
These imply that $P_{11} > 0$, hence $P_{12} = P_{21} > 0$. Thus  LQR  is a full state feedback controller.
Different from the previous subsection, multi-species systems with full-state feedback often face partial observability challenges, the following part addresses the design of biological LQR controllers under different observation scenarios.

 If $X_1$ is unmeasurable, namely,  $y=x_2(t)$, the $u_{\text{To}}$ can not be determined by the output. Since  (\ref{eq:ds}) together with output $y$ is observable, we can design an additional reduced-order state observer to estimate the concentration of  
$cX_1$ for any constant $c>0$, enabling the synthesis of a LQR.
Firstly, utilizing the measurability of $\dot{x}_2=\dot{y}$, the state variable $x_1(t)$
  can be expressed in terms of the output as $x_1=\frac{\dot{y}+\gamma_2y}{k}$.
And the associated state observer takes the form
\begin{equation}\label{eq:qs1}
\begin{split}
c\dot{\tilde{x}}_1
    &=cu-c\gamma_1\tilde{x}_1+L(\tilde{x}_1-\frac{\dot{y}+\gamma_2y}{k})
    \end{split}
\end{equation}
where the observer gain $L$ is a tunable parameter that can be selected to meet desired performance. Further, to eliminate $\dot{y}$ in $c\dot{x}$, we
apply the change of variables $\hat{x}=c\tilde{x}_1+\frac{Ly}{k}$. And the equivalent system dynamics in terms of $\hat{x}$
follows
\begin{equation}\label{eq:hatx}
    \dot{\hat{x}}=cu-(\gamma_1-\frac{L}{c})\hat{x}+(\frac{\gamma_1L}{k}-\frac{L^2}{ck}-\frac{L\gamma_2}{k})x_2.
\end{equation}
 Based on the observer variable $\hat{x}$ and output $x_2$, the state-feedback controller $u_{\text{To}}$ can be determined as  $$u_{\text{To}}=-\frac{P_{11}}{c}(\hat{x}-\bar{\hat{x}})-(P_{12}-\frac{P_{11}L}{ck})(x_2-\bar{x}_{\text{sp}})+\bar{u}$$
where $\bar{\hat{x}}=c\bar{x}_1+\frac{L\bar{x}_{\text{sp}}}{k}$. 
Substituting 
$u_{\text{To}}$
  into (\ref{eq:hatx}), the closed-loop dynamics for $\hat{x}$ is 
\begin{equation}\label{eq:hatx1}
\begin{split}
     \dot{\hat{x}}&=-(P_{11}/c+\gamma_1-L/c)(\hat{x}-\bar{\hat{x}})\\
&+\left[\frac{L(P_{11}+c\gamma_1-L)}{ck}-P_{12}-\frac{L\gamma_2}{k}\right](x_2-\bar{x}_{\text{sp}})
\end{split}
\end{equation}
The tunable observer gain $L$  can be chosen sufficiently small such that $P_{11}+c\gamma_1-L>0$ and $\frac{L(P_{11}+c\gamma_1-L)}{ck}-\frac{L\gamma_2}{k}<P_{12}$, thereby guarantees the stability of the closed-loop system.  Thus under the LQR framework, the theoretically  dynamics of $X_1$
 is given by:
\begin{equation}\label{eq:x1}
    \dot{x}_1=-\frac{P_{11}}{c}(\hat{x}-\bar{\hat{x}})-(P_{12}-\frac{P_{11}L}{ck})(x_2-\bar{x}_{\text{sp}})-\gamma_1(x_1-\bar{x}_1).
\end{equation}
The reduced observer in (\ref{eq:hatx1}) and the controller in (\ref{eq:x1})  system can be implemented by the biochemical process in Fig.\ref{fig2c}. The dynamics of the closed-loop system is
\begin{equation}\label{eq:rf}
    \begin{split}
\dot{x}_1=u_{\text{bo}}(t)-\gamma_1x_1&\triangleq h_3(x_2,\hat{x})-\gamma_1 x_1;\\~~\dot{x}_2=kx_1-\gamma_2x_2;&~~
\dot{\hat{x}}=h_4(x_2)-\gamma_{\hat{X}}\hat{x}
    \end{split}
\end{equation}
where $h_3(x_2,\hat{x})=\frac{k_{31}}{1+\frac{x_2}{K_{31}}+\frac{\hat{x}}{K_{32}}}$, $h_4(x_2)=\frac{k_{4}}{1+\frac{x_2}{K_{4}}}$ are Hill-type functions capturing the nonlinear repression and $\gamma_{\hat{X}}$ is the degradation rate of species $\hat{X}$. The dynamics after linearization at the equilibrium is
\begin{equation}
    \begin{split}\label{eq:13}
    \dot{x}_1&=h_{31}(x_2-\bar{x}_{\text{sp}})+h_{32}(\hat{x}-\bar{\hat{x}})-\gamma_1(x_1-\bar{x}_1)\\
    \dot{x}_2&=k(x_1-\bar{x}_1)-\gamma_2(x_2-\bar{x}_{\text{sp}})\\
    \dot{\hat{x}}&=h_{41}(x_2-\bar{x}_{\text{sp}})-\gamma_{\hat{X}}(\hat{x}-\bar{\hat{x}})
    \end{split}
\end{equation}
Here, $h_{31}=-\frac{k_{31}}{K_{31}\left(1+\frac{\bar{x}_{\text{sp}}}{K_{31}}+\frac{\bar{\hat{x}}}{K_{32}}\right)^2}$, $h_{32} = -\frac{k_{31}}{K_{32}\left(1+\frac{\bar{x}_{\text{sp}}}{K_{31}}+\frac{\bar{\hat{x}}}{K_{32}}\right)^2}$ and $h_{41}=  - \frac{k_{4}}{K_{4}\left(1+\frac{\bar{x}_{\text{sp}}}{K_{4}}\right)^2}$. By tuning parameters $k_{31}, k_4, K_{31}$,
$K_{32}, K_{4}$ such that $h_{31}=-P_{12}-\frac{P_{11}L}{ck}$, $h_{32}=-P_{11}/c$, $h_{41}=\frac{L(P_{11}+c\gamma_1-L)}{ck}-P_{12}-\frac{L\gamma_2}{k}$, the system (\ref{eq:rf})  can closely approximate the functionality of the theoretically LQR controller. And we can find that designing an observer for a scaled state $cX_1$, rather than $X_1$, creates more possibilities for finding a viable bio-observer $\hat{X}$. 

We now shift our perspective to the case where only $X_1$
  can be measured. Different from the previous case, the system is no longer observable. To overcome this limitation, we introduce an auxiliary species $\hat{X}_2$, which follows the same dynamics as 
$c_1X_2$ for some constant $c_1$, to estimate its value. The resulting closed-loop system is given by
\begin{equation}
    \begin{split}
   \dot{x}_1=u_{\text{bo}}(t)-\gamma_1x_1\triangleq& h_5(x_1,\hat{x}_2)-\gamma_1x_1;\\
\dot{x}_2= -\gamma_2x_2+kx_1;& ~~ \dot{\hat{x}}_2=-c_1\gamma_2 \hat{x}_2+c_1kx_1
    \end{split}
\end{equation}
where $h_5(x_1,\hat{x}_2)=\frac{k_{51}}{1+\frac{x_1}{K_{51}}+\frac{x_2}{K_{52}}}$. By appropriately tuning the parameters $k_{51}$, $K_{51}$ and $K_{52}$, the linearized dynamics around the equilibrium point $(\bar{x}_1,\bar{x}_{\text{sp}})$ can serve as an approximation of the LQR-based closed-loop system.

\textbf{Case II: Actuation via species $X_2$.} In this example, the control input is applied to species $X_2$. The dynamics of this closed loop system is 
\begin{equation}\label{eq:ds2}
\begin{split}
\dot{x}_1=\mu-\gamma_{1}x_1;~~ 
\dot{x}_2=kx_1-\gamma_2x_2+u.
\end{split}
\end{equation}
The optimal regulator in LQR framework is  $u_{\text{To}}=-(0,1)P\tilde{x}+\bar{u}$ where $P$ satisfies the Riccati equation 
\begin{equation}
 \begin{split} P_{22}^2+2\gamma_2P_{22}-Q_{22}&=0\\
 -(\gamma_1+\gamma_2)P_{21}+kP_{22}-P_{21}P_{22}&=0.
\end{split}
\end{equation}
The above equation yields $P_{22}=-\gamma_{2}+\sqrt{\gamma_2^2+Q_{22}}>0$ and $P_{21}=\frac{kP_{22}}{(\gamma_1+\gamma_2)+P_{22}}>0$. And the corresponding
$u_{\text{To}}$ is $$u_{\text{To}}=-P_{21}(x_1-\bar{x}_1)-P_{22}(x_2-\bar{x}_2)+\bar{u}.$$ The following part focuses on the biochemical  implementation of $u_{\text{To}}$ under the partial observation. 

\begin{figure}[h]	 \begin{subfigure}{.24\textwidth}		\centering  		\includegraphics[width=0.75\textwidth]{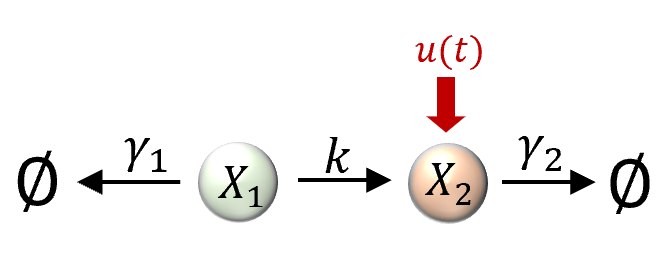}	 	\caption{open loop}\label{fig3a}
   \end{subfigure}
     \begin{subfigure}{.24\textwidth}    \centering\includegraphics[width=0.75\textwidth]{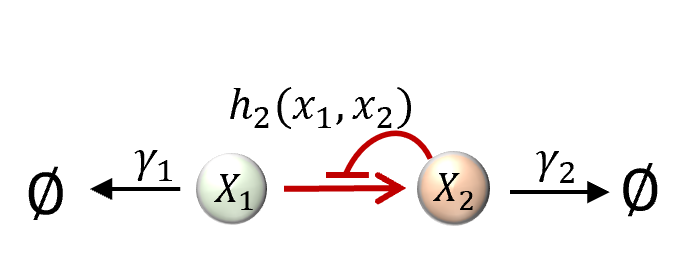}
\caption{All states are measurable.}\label{fig3b}
\end{subfigure}
      \begin{subfigure}{.24\textwidth}		\centering  		\includegraphics[width=0.75\textwidth]{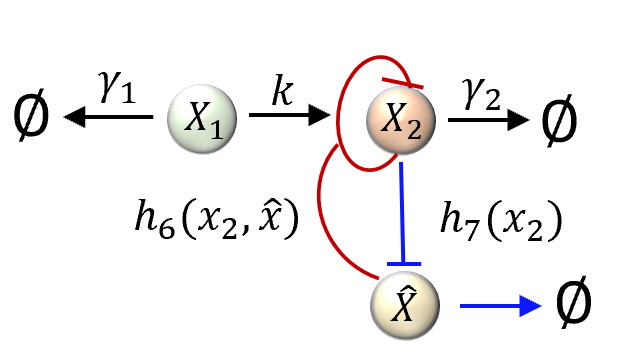}	 \caption{$X_2$ is measurable.}\label{fig3c}
   \end{subfigure}
     \begin{subfigure}{.24\textwidth}		\centering  		\includegraphics[width=0.75\textwidth]{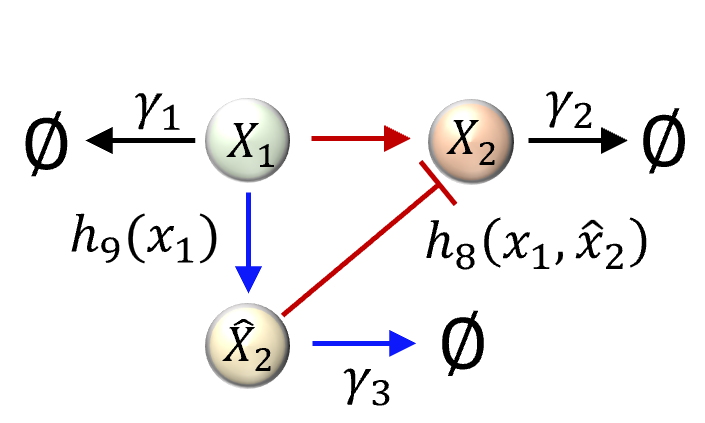}	 \caption{$X_1$ is measurable.}\label{fig3d}
   \end{subfigure}
\caption{LQR of Case II for two-gene expression process.}
     \label{fig.3}
\end{figure}

If $X_1$ is unmeasurable, that is, $y=x_2(t)$, the corresponding reduced-order state observer can be constructed due to the observability of the system. By applying the measurability of $\dot{x}_2=\dot{y}$, the observer of $c_2X_1$ for any constant $c_2>0$ is  \begin{equation}
     c_2\dot{\tilde{x}}_1=c_2\mu-c_2\gamma_1x_1+L(x_1-\frac{\dot{x}_2+\gamma_2x_2-u}{k})
  \end{equation}
Denote $\hat{x}$ as $\hat{x}=c_2x_1+\frac{Lx_2}{k}$, the dynamics of $\hat{x}$ becomes
  \begin{equation}
   \dot{\hat{x}}= c_2\mu-(\gamma_1-\frac{L}{c_2})\hat{x}+(\frac{L\gamma_1}{k}-\frac{L^2}{kc_2}-\frac{L\gamma_2}{k})x_2+\frac{Lu}{k}.  
  \end{equation}
 Thus
$u_{\text{To}}$ can be determined via the above variable substitution as follows
\begin{equation} u_{\text{To}}=-\frac{P_{21}}{c_2}(\hat{x}-\bar{\hat{x}})-(P_{22}-\frac{P_{21}L}{kc_2})(x_2-\bar{x}_{\text{sp}})+\bar{u}.
\end{equation}
Thus, the  whole closed-loop system with the reduced-order observer is described by 
\begin{equation}
\begin{split}
\dot{x}_1&=\mu-\gamma_1x_1\\
 \dot{x}_2&=k(x_1-\bar{x}_{1})-(\gamma_2+P_{22}-\frac{P_{21}L}{kc_2})(x_2-\bar{x}_{\text{sp}})-\frac{P_{21}}{c_2}(\hat{x}-\bar{\hat{x}})\\
 \dot{\hat{x}}&=-(\gamma_1-\frac{L}{c_2}+\frac{LP_{21}}{kc_2})(\hat{x}-\bar{\hat{x}})\\&+\frac{L}{k}(\gamma_1-\frac{L}{c_2}-\gamma_2-P_{22}+\frac{LP_{21}}{c_2k})(x_2-\bar{x}_{\text{sp}}).
 \end{split}
\end{equation}
From the value of $P_{21}$ and $P_{22}$,  the inequalities $(\gamma_1-\frac{L}{c_2}+\frac{LP_{21}}{kc_2}) > 0$ and $(\gamma_1-\frac{L}{c_2}-\gamma_2-P_{22}+\frac{LP_{21}}{kc_2})< 0$ hold if $L/c_2$ satisfies $\gamma_1-\gamma_2-\frac{Q_{22}}{\gamma_1+\gamma_2} \leq \frac{L}{c_2}\leq \gamma_1+\frac{\sqrt{\gamma_2^2+Q_{22}}-\gamma_2}{\gamma_1+\gamma_2}$, which ensures the stabilization of the closed-loop system. It is obvious that we can find suitable $L$ and $c_2$ and Fig.\ref{fig3c}  further shows the biological implementation of the closed-loop system, whose dynamics is
\begin{equation}\label{eq:3c}
    \begin{split}      \dot{x}_1=\mu-\gamma_1x_1;&~\dot{\hat{x}}=h_7(x_2)-\gamma_{\hat{X}}\hat{x}\\
\dot{x}_2=kx_1+u_{\text{bo}}(t)&\triangleq kx_1+h_6(x_2,\hat{x})
    \end{split}
\end{equation}
Here, the nonlinear feedback functions are defined as $h_6(x_2,\hat{x})=\frac{k_{61}}{1+\frac{x_2}{K_{61}}+\frac{\hat{x}}{K_{62}}}$ and $h_7(x_2)=\frac{k_{71}}{1+\frac{x}{K_{71}}}$. Analogous to the preceding case, the LQR and observer functions can be approximated linearly by (\ref{eq:3c}) through appropriate parameter tuning.

However, when $x_2(t)$ is not measurable, the system  (\ref{eq:ds2}) becomes unobservable. To address this, we introduce an auxiliary species $\hat{X}_2$ that mimics $c_2\dot{x}_2$, thereby providing an estimate of the unmeasured $c_3x_2(t)$ for any constant $c_3$ and enabling effective control. This construction corresponds to a well-known incoherent feedforward loop, as illustrated in Fig.~\ref{fig3d}, whose dynamics is
\begin{equation}\label{eq:dfg}
\begin{split}
\dot{x}_1=\mu-\gamma_1 x_1; ~~\dot{\hat{x}}_2&=h_{9}(x_1)-\gamma_3 \hat{x}_2\\
\dot{x}_2=u_{\text{bo}}(t)+k_1x_1-\gamma_2&x_2\triangleq h_8(x_1,\hat{x}_2)-\gamma_2x_2.
\end{split}
\end{equation} 
Here,  hill function $h_8(x_1,\hat{x}_2)=\frac{k_{81}x_1}{1+\frac{x_1}{K_{81}}+\frac{\hat{x}_2}{K_{82}}}$ captures the competitive activation by $X_1$ and inhibition by $\hat{X}_2$ on the expression of $X_2$, while $h_9(x_1)=\frac{k_9x_1}{K_9+x_1}$ describes the activation by $X_1$ on the expression of $\hat{X}_2$. 
And the linearized dynamics of $X_2$, $\hat{X}_2$ around the equilibrium $(\bar{x}_1,\bar{x}_2,\bar{\hat{x}}_2)=(\bar{x}_1,\bar{x}_{\text{sp}},c_3\bar{x}_{\text{sp}})$ is
\begin{equation}\label{eq:ds1}
\begin{split}
\dot{x}_2&=-\gamma_2(x_2-\bar{x}_{\text{sp}})+h_{81}(x_1-\bar{x}_1)+h_{82}(\hat{x}_2-c_3\bar{x}_{\text{sp}})\\
\dot{\hat{x}}_2&=h_{91}(x_1-\bar{x}_1)-\gamma_3(\hat{x}_2-c_3\bar{x}_{\text{sp}})
\end{split}
\end{equation}
where $h_{81}= \frac{k_{81} \left(1 + \frac{c_3\bar{x}_{\text{sp}}}{K_{82}} \right)}{\left(1 + \frac{\bar{x}_1}{K_{81}} + \frac{c_3\bar{x}_{\text{sp}}}{K_{82}} \right)^2}$, $h_{82}= \frac{-k_{81} \bar{x}_1}{K_{82} \left(1 + \frac{\bar{x}_1}{K_{81}} + \frac{c_3\bar{x}_{\text{sp}}}{K_{82}} \right)^2}$ and $h_{91}=\frac{k_9K_9}{(K_9+\bar{x}_1)^2}$. By tuning the parameters such that
\begin{equation}
\begin{split}\label{eq:32}
h_{91}=c_3h_{81};&~~\gamma_3=c_3(\gamma_2-h_{82})\\
h_8(\bar{x}_1,\bar{x}_{\text{sp}}) = \gamma_2 \bar{x}_{\text{sp}},~ &h_{81} = -P_{21} + k_1, ~h_{82} = -P_{22},
\end{split}    
\end{equation}
the dynamics of $\hat{X}_2$ with the parameters in the first line 
 matches the dynamics of $c_3X_2$ and can thus be interpreted as an estimation of $c_3x_2$.
Further, the estimated value $\hat{x}_2/c_3$ can be used to replace the unknown $x_2$ in the expression of $u_{\text{To}}$ in $\dot{x}_2$, thereby allowing $u_{\text{To}}$ to be fully determined.
 Thus we can conclude that the IFFL motif in (\ref{eq:dfg}) with $u_{\text{bo}}=h_8(x_1,\hat{x}_2)$ can realize the optimal control function of LQR. 

\begin{remark}
     Note that if both $X_1$ and $X_2$ can be observed, the closed-loop implementations can be constructed similarly. It is easy to verify that  LQR can be realized by Hill functions $u_{1}=h_1(x_1,x_2)=\frac{k_{11}}{1+\frac{x_1}{K_{11}}+\frac{x_2}{K_{12}}}$ in Case I (Fig. \ref{fig2b}) and $u_{2}=h_2(x_1,x_2)-kx=\frac{k_{2}x_1}{K_2+x_1+x_2}-kx$ in Case II (Fig. \ref{fig3b}) under appropriate parameters.
\end{remark}
\begin{remark}
   When applying LQR $u_{\text{To}}$ to the design of biological systems, a fundamental reality must be acknowledged: it is difficult to precisely tune biochemical parameters in the controller to their theoretical values. The effectively implemented controller operates as $\tilde{u}_{\text{To}}=-K'\tilde{x}$ where $K'\approx K$. However, this is not a fundamental flaw. Thanks to the continuous dependence of the closed-loop system on its parameters, minor deviations in 
$K$ result only in small and continuous deviations of the system trajectory from the ideal path. More critically, the inherent robust stability of the LQR controller itself ensures that these deviations are confined to an acceptable performance range, thereby guaranteeing system stability and basic functionality. This engineering characteristic aligns remarkably well with universal phenomena in biological organisms that living systems do not strive to maintain species concentrations or reaction rates at a fixed value but rather regulate them within a reasonable, dynamic range. The fundamental objective is to intelligently balance various competing demands under internal constraints of limited resources and a changing external environment, ultimately achieving a global balance between cost and benefit.  
\end{remark}
\section{Numerical Simulation}\label{sec:4}
\begin{example}
  Consider the closed loop gene expression process shown in Fig.\ref{fig2c} where $k=\gamma_1=\gamma_2=\bar{x}_{\text{sp}}=1$, observer gain $L=0.197$ and weighted matrix $Q=[q_{ij}]\in\mathbb{R}^{2\times 2}$ where $q_{22}=1$ and other elements are all zero.  
   Based on these, the LQR can be determined as 
$$u_{\text{opt}}=-0.197(x_1-1)-0.217(x_2-1).$$ 
Further solving the equations of requirements of $h_{31}, h_{32}, h_{41}$ established in Case I, we can derive that if $k_{31}=1.9670$,  $K_{31}=1.9873$, $K_{32}=2.5806$, $k_4=1.8270$ and $K_{4}=1.9012$, the hill function $h_3(x_2,\hat{x})$ can linearly approximate the LQR and $h_4(x_2)$ can make $\hat{x}$ serve as an observer. 

Comparison of the optimal closed-loop control system (\ref{eq:rf}) and the open-loop control system (\ref{eq:ds1}) with $u=1$ is conducted from the following simulation. The first subplot illustrates the concentration evolution of the species in LQR closed-loop ($X_1$, $X_2$ and $\hat{X}$ correspond to the gray, red, and dark gray line, respectively.) and open-loop control strategies ($X_1$, $X_2$ corresponds to gray, red dashed line). The initial state is $(x_1(0),x_2(0),\hat{x}(0))=(0.5, 4, 2.5)$ and the final convergent state is $(1,1,1.197)$. Two disturbances are applied: the first from 4s to 6s with a magnitude of 5, and the second from 15s to 17s with a magnitude of –1. The figure highlights the system trajectories and the convergence times (defined as the time when the error remains within ±5\% of the reference value), with blue lines representing the closed-loop and orange lines representing the open-loop responses. It is evident that the closed-loop system achieves faster convergence both from the initial state and after each disturbance.
The second subplot reflects the magnitude of control input $u(t)$. The green line and red dashed
 line reflect the evolution of control input in closed-loop control and open-loop control, respectively. The last subplot presents the evolution of instantaneous cost and the  accumulated cost values of three orange regions. The first region spans from 0s to 3.5s, during which the cost function values for the open-loop and closed-loop systems are 
$J=3.95$ and 
$J=3.84$, respectively. The second region covers the interval from 5s to 13s, with corresponding cost values of 
$J=
20.52$ (open loop) and 
$J=17.55$ (closed loop). The third region extends from 15s to 22s, where the cost function values are 
$J=0.84$ for the open-loop case and 
$J=0.69$ for the closed-loop case. These three figures shows that the concentration of the target species 
$x_2(t)$ exhibits faster convergence, stronger disturbance rejection, and 
lower control cost with the presence of  the biomolecular LQR. 
  \begin{figure}[h]
       \centering \includegraphics[width=1\linewidth]{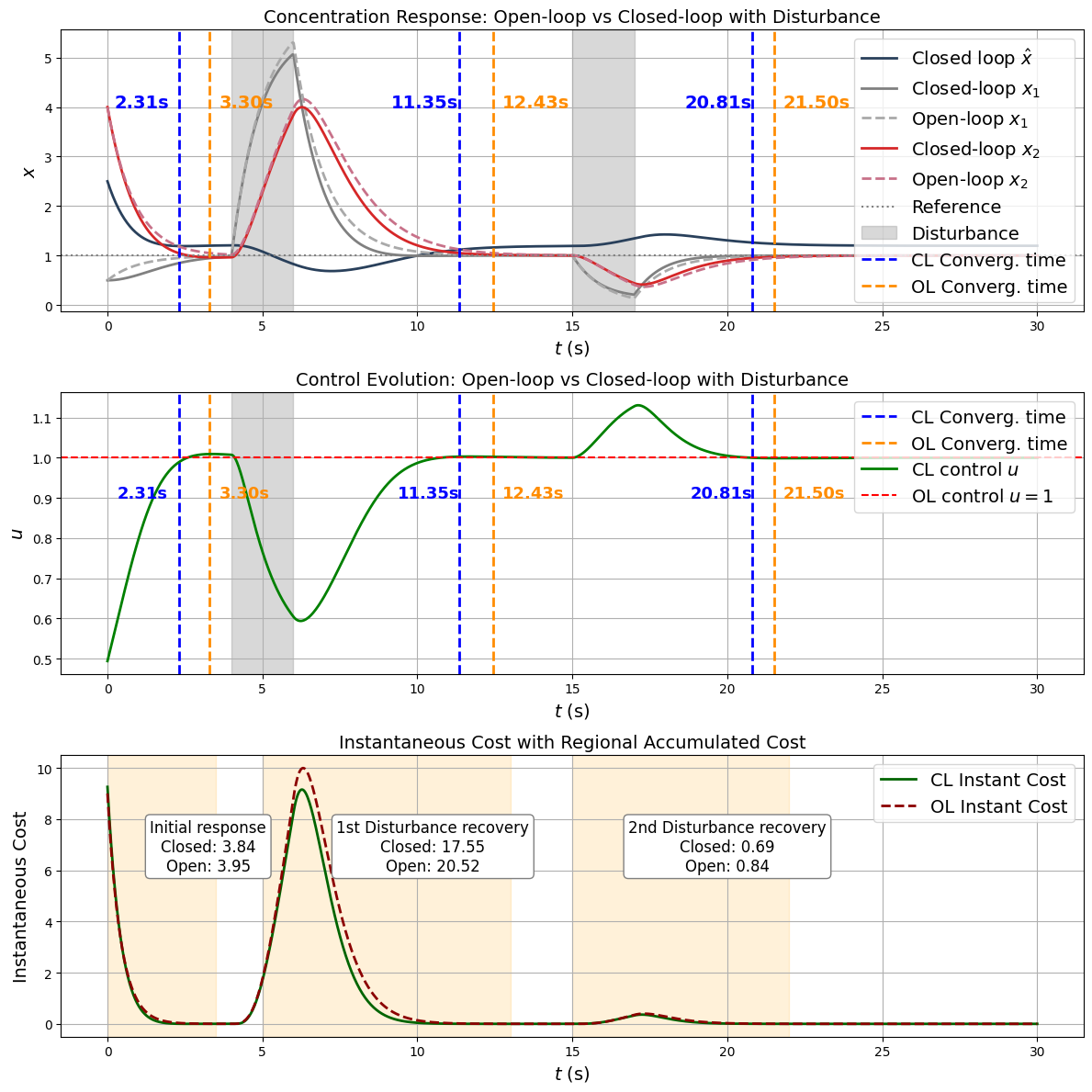}
       \caption{Camparsion between closed loop and open loop.}
       \label{fig4a}
   \end{figure}
  \begin{example}
We now shift our focus to the closed-loop system in Case II, as illustrated in Fig.~\ref{fig3d}, where $X_2$ is not directly measurable. The parameters in the corresponding dynamics~(\ref{eq:dfg}) are set to $k = \gamma_1 = \gamma_2 = \bar{x}_{\text{sp}} = 1$, and the weighting matrix $Q$ shares the same definition as that in the previous example. 
The solution to the Riccati equation under this configuration yields $P_{12} = 3 - 2\sqrt{2}$ and $P_{22} = \sqrt{2} - 1$, resulting in the following LQR controller:
\begin{equation}
u_{\text{opt}}=-(3-2\sqrt{2})(x_1-1)-(\sqrt{2}-1)(x_2-1). 
\end{equation}
By selecting $k_{81} = 1.39$, $K_{81} = 4.19$, $K_{82} = 6.52$, $k_9 = 3.41$, and $K_9 = 1.41$, the system~(\ref{eq:dfg}) is capable of approximating both the estimation and control behaviors specified by the LQR solution.

Figure~\ref{fig4a} compares the dynamic responses of the open-loop system (i.e., system~(\ref{eq:ds2}) with $u = 0$) and the closed-loop system~(\ref{eq:dfg}) under two distinct disturbances. The first disturbance occurs between 10~s and 12~s with a magnitude of 3, and the second between 22~s and 24~s with a magnitude of $-1$. For consistency, the graphical representations in this example follow the same plotting conventions as in the previous example. The line styles and color codes retain the same meanings.
   The first subplot illustrates the evolution of the concentrations of three species from the initial state $x(0) = (5,\ 5,\ 2.5)$. Since $X_1$ is uncontrollable, its trajectory (shown in gray) is identical in both the open- and closed-loop scenarios. 
  Although the control input in the closed-loop case (shown in Case II) exhibits a larger absolute value than in the open-loop case (where $u \equiv 0$), the cost shown in the third subplot is consistently higher in the open-loop system. We also computed the accumulated costs over three orange-shaded time intervals: (1) 0–7s, with costs of 15.55 (closed loop) and 20.32 (open loop); (2) 10–17s, with 4.43 and 7.54, respectively; and (3) 22–29s, with 0.68 and 0.84. These results show that the closed-loop design improves disturbance rejection, accelerates convergence, and reduces overall cost.
\end{example}

    \begin{figure}[h]
       \centering \includegraphics[width=1\linewidth]{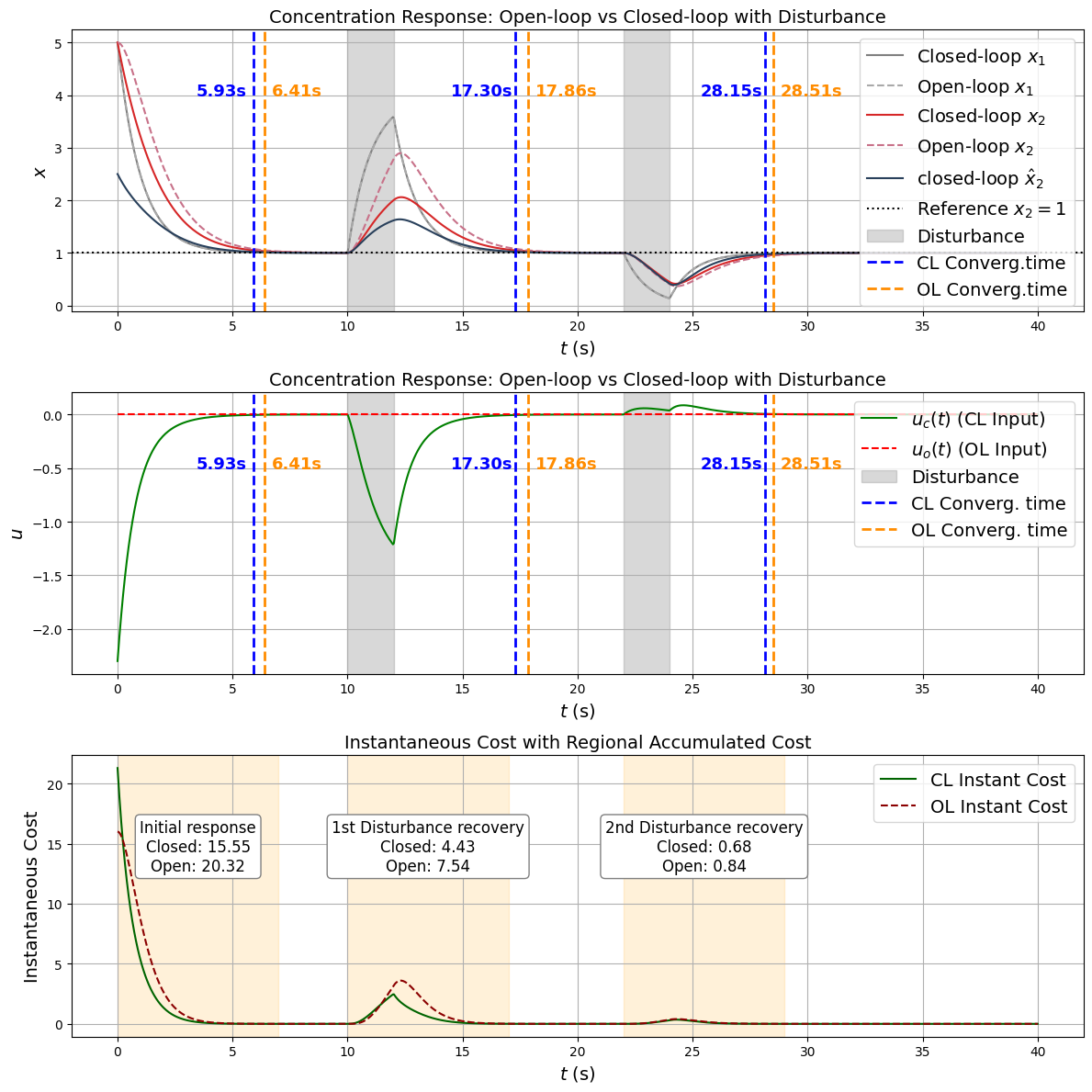}
\caption{Comparison between closed loop and open loop.}
       \label{fig4c}
   \end{figure}
\end{example}

\section{Conclusion and Discussion}\label{sec:5}
This paper proposes a biochemical Linear Quadratic Regulator (LQR) framework for the regulation of one- and two-gene expression processes under partial observability. We show that the derived biomolecular LQR controllers structurally align with commonly observed network motifs in biological systems. This alignment not only offers a functional rationale for their prevalence but also supports the view that biological circuits implement control strategies that effectively balance performance with cellular cost.

Furthermore, this framework provides a powerful explanatory lens for understanding the design principles of living systems. In the profound correspondence between LQR and biological organisms, the Q matrix is not merely an abstract mathematical setting; it is shaped by the fundamental survival goals of the organism and the specific constraints of its environment. This evolutionarily emergent $Q$ matrix provides a directional blueprint: it drives the formation of specific network topologies (e.g., explaining why a certain negative feedback loop, IFFL loop exists) and delineates the adjustment range for key parameters (e.g., defining the requisite feedback strength), thereby guiding system behavior to serve fundamental survival value. Consequently, the significance of LQR extends beyond a design tool. It also serves as a powerful explanatory framework that reveals how natural selection, acting as a dynamic optimization process, shapes the complex and robust regulatory systems we observe through feedback.

In future work, we aim to extend the bio-LQR framework to more general classes of biochemical processes, with a specific focus on incorporating LQI principles which is essential for asymptotically rejecting persistent disturbances and eliminating steady-state errors, thereby achieving a higher degree of regulatory precision. Two major challenges remain: (1) understanding the relationship between the solution of the Riccati equation and the underlying network structure of the biochemical system, and (2) developing appropriate model reduction techniques for both the controller and the observer in large-scale biochemical systems.


\appendix

\end{document}